# Non-exponential tunneling ionization of atoms by an intense laser field

**A.M. Ishkhanyan**[1] **and V.P. Krainov**[2]

[1]Institute for Physical Research, NAS of Armenia, 0203 Ashtarak, Armenia
[2]Moscow Institute of Physics and Technology, 141700 Dolgoprudny, Russia

We discuss the possibility of non-exponential tunneling ionization of atoms irradiated by intense laser field. This effect can occur at times, which are greater than the lifetime of a system under consideration. The mechanism for non-exponential depletion of an initial quasi-stationary state is the cutting of the energy spectrum of final continuous states at long times. We first consider the known examples of cold emission of electrons from metal, tunneling alpha-decay of atomic nuclei, spontaneous decay in two-level systems and the single-photon atomic ionization by a weak electromagnetic field. The new physical situation discussed is tunneling ionization of atoms by a strong low-frequency electromagnetic field. In this case the decay obeys $\sim 1/t$ power-law dependence on the (long) interaction times.

## 1. Introduction

Non-exponential decay in various quantum systems has been a subject of many discussions already for a long time starting from the pioneering work by Khalfin [1]. The interest here is of conceptual character since the pure exponential decay law is not fully consistent with quantum mechanics [2-4] though this law is a somewhat universal hallmark of unstable systems in many fields of science [3]. Deviations from the exponential law were predicted for both short and long times [1-22].

Khalfin has shown that all states that have a lowest energy in their spectrum eventually, at long times, must decay more slowly than exponentially. The following discussions revealed a number of possible mechanisms for non-exponential decay for long as well as short times (see, e.g., [2-22]). Different physical situations have been considered starting from a basic model of penetration through a delta-edged potential [5], and including such processes as the spontaneous decay in two-level systems [6], tunneling alpha-decay of atomic nuclei and cold electron emission from metals [7], single photon ionization of atoms [8], non-exponential decays in autoionizing states [9-10], decay of excited helium state [11], etc. Among these mechanisms one should mention frequent measurements which slow the evolution of a quantum system, hindering transitions to states different from the initial one.

This phenomenon, known as the quantum Zeno effect may lead to infinitely slow decay [12]. Different phenomena where the decay is not at all exponential are compared in several monographs (e.g., [13), and reviews (e.g., [2]) including recent ones [14,15].

Power-law time-dependences of the type $t^{-n}$ for non-exponential decay at long times have been revealed for a number of situations. For instance, analyzing the quantum non-exponential decay in many-particle systems, it was found that a $t^{-N}$ time dependence is realized at long times where the quantity $N$ is proportional to the number of particles and depends on the quantum statistics of these particles, Bose or Fermi [16]. The exponential and non-exponential regimes of the buildup process in resonant tunneling structures are studied in a recent paper [17] by considering an analytic solution of the time-dependent Schrödinger equation. It was found that the probability amplitude exhibits a purely exponential behavior in a finite time interval followed by a clear transition to a non-exponential regime. The probability amplitude in the non-exponential regime follows a $t^{-3/2}$ time dependence. Several possibilities for different power-laws for long times are suggested by the decay of autoionizing states. A possible $\sim 1/t$-dependence of decay amplitude has been mentioned in [9], however, the more rigorous calculations have shown that for the discussed physical context of ionization more accurate is the $\sim 1/t^{3/2}$ estimation [10]. The decay is exponential when the energy density distribution of states is of Lorentzian form. Thought this is common for open systems, in isolated interacting quantum systems, various deviations from the Lorentzian shape, leading to non-exponential decays, may occur [18]. For instance, if for the energy density distribution a Lorentzian profile containing an additional threshold factor is considered, the result is the dominant logarithmic decay $1/(t\log^2 t)$ over long times [19].

Deviations from the exponential law have been observed experimentally (see, e.g., [4,20,21]), and the difficulties for the experimental verification of the non-exponential decay including the ones related to fluctuations have been discussed (see, e.g., [3,5]). Difficulties for the experimental verification of long-time non-exponential decay include, along with the weakness of the decaying signal, the measurement itself, because of the suppression of the initial state reconstruction [22]. Our result for the non-resonant tunneling ionization of atoms by a strong low-frequency electromagnetic field is that the decay at long times exhibits a $t^{-1}$ time dependence. It is understood that this is advantageous for a possible experimental observation because of slow suppression of the initial state. A promising process for observation is the near-threshold photo-detachment of electrons from negative ions.

## 1. Spontaneous decay in a two-level system

Exponential decay of atomic systems in weak external fields is a well known phenomenon [23]. The simplest example is the spontaneous decay in a two-level system (Fig. 1). A charged particle is assumed to be initially ($t = 0$) in the upper state 1 with the energy $E_1$. The particle emits a spontaneous photon with the frequency $\omega$ close to $E_1$ (here and hereafter we put $\hbar = 1$) and goes to the ground state 0 with the energy $E_0 = 0$.

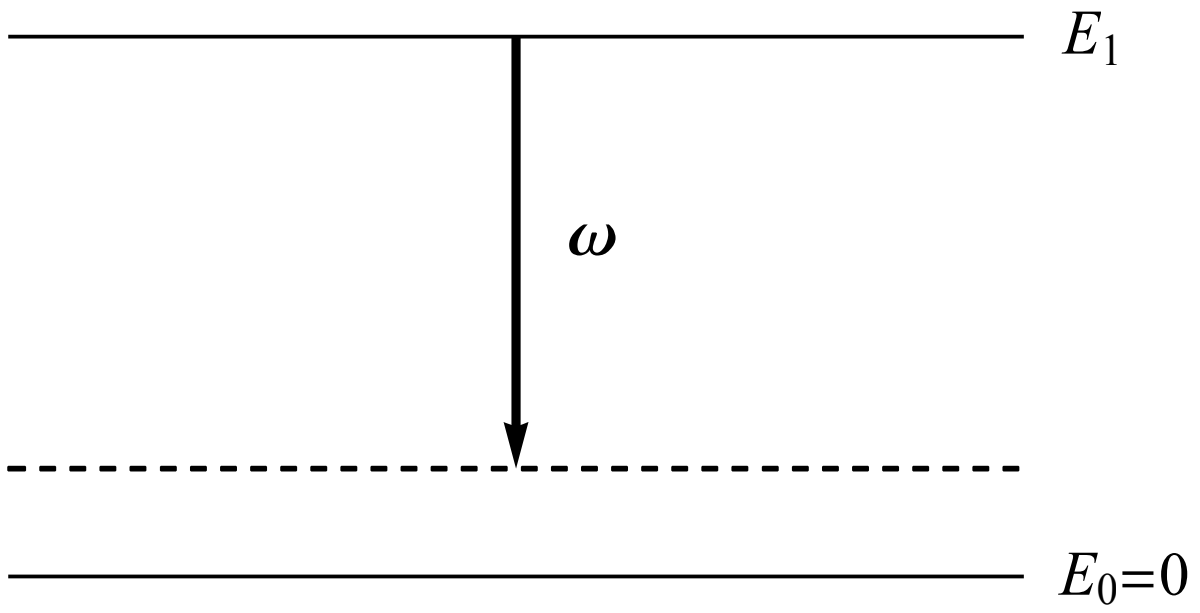

Fig. 1. Spontaneous decay in a two-level system.

According to Fermi's golden rule the transition rate for transitions with the frequency of emitted photon close to $E_1$ is given by [24]

$$\Gamma(E_1) = 2\pi \int \left|V_{10}\right|^2 \delta(E_1 - \omega) \frac{2\omega^2 d\omega\, d\Omega}{(2\pi c)^3}. \tag{1}$$

It is assumed that the levels are not close each to other, so that $\Gamma(E_1) << E_1$, hence, a weak interaction with the field of the electromagnetic vacuum takes place. Here $\Omega$ is the solid angle, $V_{10}$ is the (non-relativistic) dipole matrix element, and the interaction with the field of the electromagnetic vacuum is given as

$$V e^{i\omega t} = \sqrt{\frac{2\pi}{\omega}} \vec{P} \vec{e}\, e^{i\omega t}, \tag{2}$$

where $\vec{P}$ is the momentum operator, $\vec{e}$ is the polarization of the emitted photon; $e = m = 1$ here and thereafter. The normalization volume is assumed to be equal to unity (it disappears in physical results). Then the absolute probability for the transition $1 \to 0$ is linear in time $t$. This is correct if $t < 1/\Gamma$.

The amplitude $a_1(t)$ of the upper state at $t < 1/\Gamma$ is of the form [25] (we prove this statement later)

$$a_1(t) = e^{-iE_1 t - \Gamma(E_1)t/2}. \tag{3}$$

At $t \to \infty$, $a_1(t) \to 0$. Consider now the photon energy spectrum at the limit $t \to \infty$. We recall that the frequency $\omega$ is close to $E_1$, but is not exactly equal to $E_1$. Let $a_{0\vec{k}}(t)$ be the amplitude of the lower state if a photon with wave vector $\vec{k}$ is emitted ($k = \omega / c$). Since the interaction $V$ is small, the first-order perturbation theory can be applied to determine $a_{0\vec{k}}(t)$. Using the Schrödinger equation,

$$i \frac{da_{0\vec{k}}(t)}{dt} = \left\langle 0\vec{k} \left| V \right| 1 \right\rangle e^{i\omega t} a_1(t), \tag{4}$$

and integrating over time, one obtains the amplitude of the lower state:

$$a_{0\vec{k}}(t) = \left\langle 0\vec{k} \left| V \right| 1 \right\rangle \frac{1 - \exp[i(\omega - E_1)t - \Gamma(E_1)t/2]}{\omega - E_1 + i\Gamma(E_1)/2}. \tag{5}$$

At $t \to \infty$ the transition probability tends to

$$\left| a_{0\vec{k}}(t) \right|^2 = \frac{\left| \left\langle 0\vec{k} \left| V \right| 1 \right\rangle \right|^2}{(\omega - E_1)^2 + \Gamma^2(E_1)/4}. \tag{6}$$

Now we consider all polarizations of the emitted spontaneous photon, and integrate over the solid angle. Introducing the notation

$$\Gamma(\omega) d\omega = 2\pi \int \left| \left\langle 0\vec{k} \left| V \right| 1 \right\rangle \right|^2 d\Omega \frac{2\omega^2 d\omega}{(2\pi c)^3}, \tag{7}$$

the result is written as

$$\left| a(\omega) \right|^2 = \frac{1}{2\pi} \frac{\Gamma(\omega)}{(\omega - E_1)^2 + \Gamma^2(E_1)/4}. \tag{8}$$

Thus, the well known Breit-Wigner partial distribution is realized for the spectrum of emitted photons [25]. It follows from Eq. (8) that

$$\int_0^\infty \left| a(\omega) \right|^2 d\omega \approx 1. \tag{9}$$

This is well understood since a spontaneous photon can only be emitted; it cannot be absorbed. The main contribution to this integral comes from the region $\omega \approx E_1$ (note that Eqs. (8) and (9) are valid if $\Gamma << E_1$).

### 2.1. Fock-Krylov theorem.

Considering the limit $t \to \infty$, we introduce a system of continuous eigenfunctions $\Psi_{0\omega}(x)$ describing the final state of the lower level 0, which correspond to radiation of a spontaneous photon of frequency $\omega$. Let us first expand the wave function of the upper level $\Psi_1(x)$ in terms of these functions at $t = 0$:

$$\Psi_1(x,t=0)=\int_0^{\infty} a(\omega)\Psi_{0\omega}(x)\,dx\,. \tag{10}$$

The wave function of the system at an arbitrary time instant $t$ is written as

$$\Psi_1(x,t)=\int_0^{\infty} a(\omega)\Psi_{0\omega}(x)e^{-i\omega t}d\omega\,. \tag{11}$$

Thus, the probability amplitude $a_1(t)$ for a particle initially being at the upper state 1 is

$$a_1(t)=\int_0^{\infty}\Psi_1^*(x,t)\Psi_1(x)dx=\int_{-\infty}^{\infty}\int_0^{\infty}\int_0^{\infty} a^*(\omega)\,\Psi_{0\omega}^*(x)e^{i\omega t}a(\omega')\Psi_{0\omega'}(x)\,dx\,d\omega\,d\omega'\,. \tag{12}$$

Since
$$\int_{-\infty}^{\infty}\Psi_{0\omega}^*(x)\,\Psi_{0\omega'}(x)dx=\delta(\omega-\omega')\,, \tag{13}$$

we obtain from Eq. (12) that

$$a_1(t)=\int_0^{\infty}|a(\omega)|^2\,e^{-i\omega t}d\omega\,. \tag{14}$$

This is the Fock-Krylov theorem [26]. At long times, $t\to\infty$, the most significant contribution is from the low frequencies $\omega\approx 0$.

### 3. Exponential decay of the upper level

Let us substitute the Breit-Wigner distribution (8) into Eq. (14):

$$a_1(t)=\frac{1}{2\pi}\int_0^{\infty}\frac{e^{-i\omega t}\Gamma(\omega)}{(\omega-E_1)^2+\Gamma^2(E_1)/4}d\omega\,. \tag{15}$$

The integration contour here can be shifted down to the minus infinity (Fig. 2). Then, the contribution of the lower horizontal line can be neglected, and the simple pole at $\omega=E_1-i\Gamma(E_1)/2$ gives the contribution

$$a_1^{(1)}(t)=e^{-iE_1t-\Gamma(E_1)/2}\,. \tag{16}$$

Thus, we confirmed the above postulated Eq. (3). Note that this result is in agreement with the Wigner-Weisskopf approach to the exponential decay [25].

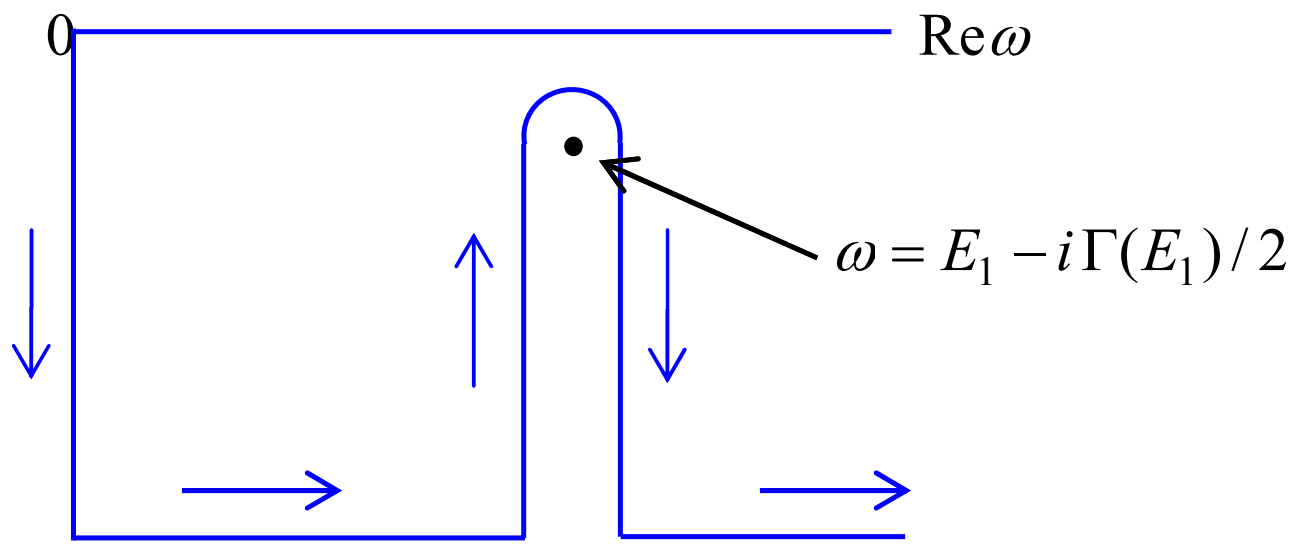

Fig. 2. The integration contour in Eq. (15).

**4. Non-exponential decay of the upper level**

In order to get a basic insight into the mechanism of formation of a non-exponential decay law, consider the following example.

The Fock-Krylov theorem is applicable if the lower integration limit in Eq. (14) is finite. This limit in our case is $\omega = 0$ since, as it was said above, spontaneous photons can be only emitted, not absorbed. One has to take into account the behavior of the amplitude on the real axis, i.e. only on physically accessible regions, avoiding the possible ambiguities associated with the poles in the complex plane [2].

Since the integration contour is deformed as shown on Fig. 1, apart from the contribution coming from the pole at $\omega = E_1 - i\Gamma(E_1)/2$, one has to take into account the contribution of the integration over the left vertical line of the contour in Fig. 2, at $\mathrm{Re}\,\omega = 0$. Introducing the notation $\omega = -iz$, Eq. (15) is rewritten as

$$a_1^{(2)}(t) = -\frac{i}{2\pi E_1^2}\int_0^{\infty} e^{-zt}\,\Gamma(-iz)\,dz\,. \tag{17}$$

Now we need the width $\Gamma(\omega)$ at $\omega \to 0$. The result slightly differs from $\Gamma(E_1)$ derived by Fermi's golden rule. After averaging of $\left|\left(\hat{\vec{P}}_{10}\vec{e}\right)\right|^2$ over the angles between $\vec{e}$ and $\vec{P}$ one finds an additional factor 1/3:

$$\left|\left\langle 0\vec{k}\right|V\left|1\right\rangle\right|^2 = \frac{2\pi}{3\omega}\left|\hat{\vec{P}}_{10}\right|^2. \tag{18}$$

Thus, for $\omega \to 0$, we obtain

$$\Gamma(\omega)d\omega = 2\pi\int \frac{2\pi}{3\omega}\left|\hat{\vec{P}}_{10}\right|^2 \frac{2\omega^2\,d\omega\,d\Omega}{(2\pi c)^3}, \tag{19}$$

i.e.

$$\Gamma(\omega) = \frac{4\omega\left|\hat{\vec{P}}_{10}\right|^2}{3c^3}. \tag{20}$$

This result reveals vanishing of the decay rate at zero-frequency limit as well as states that the radiation decay rate at a given frequency is smaller than the field oscillation frequency: $\Gamma(\omega) << \omega$. This is the physical meaning of Eq. (20), which shortly leads to non-exponential decay. Indeed, substituting this result into Eq. (17), we get

$$a_1^{(2)}(t) = -\frac{2\left|\vec{P}_{10}\right|^2}{3\pi c^3 E_1^2}\int_0^{\infty} z\,e^{-zt}dz = -\frac{2\left|\vec{P}_{10}\right|^2}{3\pi c^3 E_1^2 t^2}. \tag{21}$$

Since $\vec{P}_{10} = iE_1\vec{r}_{10}$, we finally have

$$a_1^{(2)}(t) = -\frac{2\left|\vec{r}_{10}\right|^2}{3\pi c^3 t^2}. \qquad (22)$$

Restoring usual units, we find the dimensionless amplitude of the upper state to be

$$a_1^{(2)}(t) = -\frac{2}{3\pi}\left(\frac{e^2}{\hbar c}\right)\left(\frac{\vec{r}_{10}}{c\,t}\right)^2. \qquad (23)$$

This expression is applicable for long times. A similar result was obtained in [6] by a considerably complicated approach, using resolvent perturbation theory and examining generalized Stark shifts in the complex plane.

Thus, a power-law decreasing rate of the amplitude dominates for long times such that

$$t > \frac{1}{\Gamma(E_1)}\ln\frac{c^3}{\left|\vec{r}_{10}\right|^2 E_1^2}. \qquad (24)$$

Note that in this case the rate of the spontaneous emission is given by the known expression

$$\Gamma(E_1) = \frac{2\left|\vec{r}_{10}\right|^2 E_1^3}{3c^3}. \qquad (25)$$

## 5. Single-photon ionization of a hydrogen atom

Now we consider the single-photon ionization of a hydrogen atom by a weak electromagnetic field with a fixed frequency $\omega$. Let a hydrogen atom be in its ground state at $t = 0$. Its initial energy is $-1/2$ (in atomic units) and $\omega > 1/2$. At $t \to \infty$ the atom is totally ionized, and the electron has a continuous spectrum which is again described by the Breit-Wigner distribution:

$$\left|a(E)\right|^2 = \frac{1}{2\pi}\frac{\Gamma(E)}{(\omega - 1/2 - E)^2 + \Gamma^2(\omega)/4}, \qquad (26)$$

where $E > 0$ is the kinetic energy of ejected electron. Again, this distribution is normalized:

$$\int_0^\infty \left|a(E)\right|^2 dE \approx 1. \qquad (27)$$

Since we assume $\Gamma(E) << 1$, the main contribution here is by the region $E \approx \omega - 1/2$.

Dipole interaction of an electron with the external electromagnetic field having strength $F$ is

$$V e^{-i\omega t} = \frac{\hat{\vec{P}}\vec{F}}{\omega} e^{-i\omega t}, \qquad (28)$$

where $\hat{\vec{P}}$ is the electron momentum operator. This equation describes absorption of a photon from an external field with a fixed frequency $\omega$ (unlike the case of spontaneous decay). The width $\Gamma(E)$ is then derived like in the case of spontaneous decay:

$$\Gamma(E) = \frac{2F^2 P}{3} \left| \vec{r}_{0\vec{P}} \right|^2 . \tag{29}$$

When the electron momentum $P$ goes to zero, the quantity $\left| \vec{r}_{0\vec{P}} \right|^2 \approx 128\pi / (\mathrm{e}^4 P)$ (here "e" is the base of the natural logarithm). So that at small energies for a hydrogen atom we have

$$\Gamma(E \to 0) = \Gamma_0 = \frac{256\pi F^2}{3\mathrm{e}^4} . \tag{30}$$

This is also a well known result [23].

According to the Fock-Krylov theorem the amplitude of the ground state is given as

$$a_0(t) = \int_0^\infty \left| a(E) \right|^2 e^{-iEt} dE . \tag{31}$$

Substituting Eq. (26) , we find

$$a_0(t) = \frac{1}{2\pi} \int_0^\infty \frac{\Gamma(E) \exp(-iEt) dE}{(\omega - 1/2 - E)^2 + \Gamma^2(\omega - 1/2)/4} . \tag{32}$$

The integration contour $C$ can be shifted down up to infinity (Fig. 3) analogously to the above considered case of spontaneous decay. Then the contribution of a simple pole at $E = \omega - 1/2 - i\Gamma(\omega - 1/2)/2$ determines an exponential decay:

$$a_0^{(1)}(t) = e^{it/2 - \Gamma(\omega - 1/2)t/2 - i\omega t} , \tag{33}$$

where the factor $e^{-i\omega t}$ takes into account the energy of the absorbed photon.

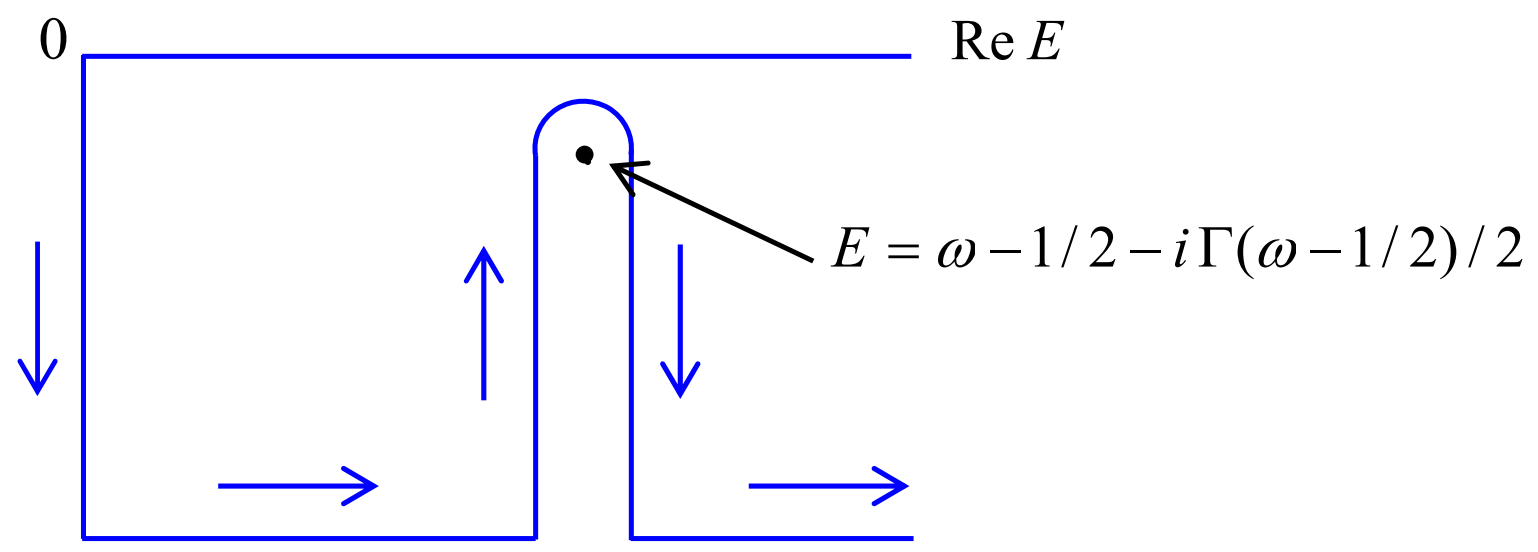

Fig. 3. Shift of the integration contour in Eq. (32).

Consider now the integral over the left vertical line in Fig. 3. Substituting $E = -iz$, we have

$$a_0^{(2)}(t) = \frac{-i\Gamma_0}{2\pi(\omega - 1/2)^2} \int_0^\infty e^{-zt} dz = \frac{-i\Gamma_0}{2\pi(\omega - 1/2)^2 t}. \tag{34}$$

Here, the finite quantity $\Gamma_0 = \Gamma(E \to 0)$ presents the rate of the ionization into the threshold $\omega = 1/2$. This equation is applicable to long times. A similar result was obtained in [8] using an approach similar to that of [6], i.e., Laplace transformation of equations for transition amplitudes and inspecting the branch cuts in the generalized Stark shifts.

If $a_0^{(2)}(t) > a_0^{(1)}(t)$, one finds that the non-exponential decay dominates at times

$$t > \frac{1}{F^2} \ln \frac{(\omega - 1/2)^2}{F^2}, \quad F << \omega - 1/2. \tag{35}$$

We neglected here the contribution from bound Rydberg states. These corrections are significant only in the region close to the edge of the continuum spectrum (see [8]), i.e. when $\omega - 1/2 << \Gamma_0$.

## 6. Non-exponential tunneling ionization of atoms by a strong low-frequency electromagnetic field

In this section we consider the tunneling ionization of atoms by a strong low-frequency electromagnetic field when the Keldysh parameter [27] is small:

$$\gamma = \frac{\omega\sqrt{2E_i}}{F} << 1. \tag{36}$$

Here $E_i$ is the ionization potential, $\omega$ and $F << F_{\text{atom}}$ are the frequency and field strength amplitude, respectively, $e = \hbar = m = 1$. Exponentially small rate of tunneling into the state with electron's longitudinal kinetic energy $E > 0$ is [28,29]

$$\Gamma(E) = 4\sqrt{\frac{3}{\pi F}} \exp\left(-\frac{2(1 - E\gamma^2)}{3F}\right). \tag{37}$$

We neglected here the transverse energy of the electron. For the sake of simplicity we consider again ionization of a hydrogen ground state. The rate (37) is averaged over the field period $T = 2\pi/\omega$.

Normalized partial distribution of the energy of emitted electrons is again of the Breit-Wigner form with respect to the mean value $\varepsilon$:

$$
|a(E)|^2 = \frac{1}{2\pi} \frac{\Gamma(E)}{(E-\varepsilon)^2 + \Gamma^2(\varepsilon)/4}. \tag{38}
$$

This electronic spectrum with one maximum is confirmed by experimental data [30] in tunneling regime at the tunnel ionization of potassium and xenon atoms by strong low-frequency $CO_2$ laser field.

According to the Fock-Krylov theorem (14), we have

$$
a_0(t) = \int_0^\infty |a(E)|^2 e^{-iEt} dE. \tag{39}
$$

The simple pole in (38) produces exponential depletion of the ground hydrogen state:

$$
a_0^{(1)}(t) = e^{-i\varepsilon t - \Gamma(\varepsilon)t/2}. \tag{40}
$$

At $t \to \infty$ the non-exponential depletion is determined by small values of $E \to 0$. According to (37), we have [27-28]

$$
\Gamma(0) = 4\sqrt{\frac{3}{\pi F}} \exp\left(-\frac{2}{3F}\right). \tag{41}
$$

Then, it follows from Eq. (39) that

$$
a_0^{(2)}(t) = -\frac{i\Gamma(0)}{2\pi\varepsilon^2 t} \tag{42}
$$

so that for the amplitude of the initial state we finally obtain

$$
a_0^{(2)}(t) = -2i\sqrt{\frac{3}{\pi^3 F}} \exp\left(-\frac{2}{3F}\right)\frac{1}{\varepsilon^2 t}. \tag{43}
$$

This result is analogous to the case of single-photon ionization of a hydrogen atom. In both cases the dependence on time as well as the dependence on the excess energy above the threshold are the same ($\sim 1/t$ and $\sim 1/\varepsilon^2$, respectively). This is a main observation of the present paper. It is understood that $\sim 1/t$ decay law is advantageous for experimental observation compared with $\sim t^{-\sigma}$, $\sigma > 1$, power-law dependence.

## 7. Discussion

As it was mentioned in introduction, numerous physical mechanisms for non-exponential decay have been discussed in past, though the experimental observations for both short [4,20] and long [20,21] time deviations are rare. For short times, the non-exponential tunneling delay was experimentally observed for the first time in [4]. In this experiment ultra-cold atoms were trapped in an accelerating periodic optical potential created by a standing wave of light. Atoms can escape the wells by quantum tunneling, and the numbers that remain can be measured as a function of interaction time. Exponential decay is modeled by the Landau-Zener tunneling process. However, non-exponential decay was observed at small times; so the explanation could be that the measured times were $<<1/\Gamma$.

Because of the suppression of the initial state reconstruction, the difficulties for the experimental verification of long-time non-exponential decay include, along with the weakness of the decaying signal, the measurement itself [5,22]. Our calculations for two physical situations, namely, for a single photon ionization of hydrogen atom and for the tunneling ionization of atoms by strong low-frequency laser field reveal $\sim 1/t$ dependence for both cases. It is understood that $\sim 1/t$ law is advantageous for a possible experimental observation in long times because of relatively slow suppression of the initial state. The results show that a promising process for experimental observation of a non-exponential decay at long times is the near-threshold photo-detachment of electrons from negative ions, since in this case the amplitude of the non-exponential decay, according to Eq. (34), is proportional to $(\omega - E_i)^{-2}$, where $E_i$ is the ionization energy.

## Acknowledgments

This research has been conducted within the scope of the International Associated Laboratory (CNRS-France & SCS-Armenia) IRMAS. The research has received funding from the European Union Seventh Framework Programme (FP7/2007-2013) under grant agreement No. 295025 – IPERA. The work has been supported by the Armenian State Committee of Science (SCS Grant No. 13RB-052) and the Russian Foundation of Basic Research (Grant No. 13-02-00072). One of the authors (V.P. K.) gives thanks for the support of the Ministry of Education and Science of Russia (State assignment No. 3.679.2014/K).